\documentclass[12pt]{iopart}

\usepackage{graphicx}
\usepackage{harvard}
\usepackage{xcolor}
\usepackage{multicol}
\usepackage{multirow}
\usepackage{subcaption}
\usepackage{bm}
\usepackage{booktabs}

\newlength\savewidth\newcommand\shline{\noalign{\global\savewidth\arrayrulewidth
  \global\arrayrulewidth 1.5pt}\hline\noalign{\global\arrayrulewidth\savewidth}}
  
\begin{document}

\title[AI-guidance improves performance of transformers]{Improving ovarian cancer segmentation accuracy with transformers through AI-guided labeling}

\author{Aneesh Rangnekar$^1$, Kevin Boehm$^2$, Emily Aherne$^{3*}$, Ines Nikolovski$^{3*}$, Natalie Gangai$^3$, Ying Liu$^4$, Dimitry Zamarin$^{4*}$, Kara Long Roche$^5$, Sohrab Shah$^2$, Yulia Lakhman$^3$ and Harini Veeraraghavan$^1$}

\address{Memorial Sloan Kettering Cancer Center}
\address{$^1$ Department of Medical Physics}
\address{$^2$ Computational Oncology, Department of Epidemiology and Biostatistics}
\address{$^3$ Department of Radiology}
\address{$^4$ Department of Medical Oncology}
\address{$^5$ Department of Surgical Oncology}
\address{$^*$ relevant work performed while at Memorial Sloan Kettering Cancer Center}
\eads{\mailto{rangnea@mskcc.org}, \mailto{veerarah@mskcc.org}}

\begin{abstract}
Transformer models have demonstrated the capability to produce highly accurate segmentation of organs and tumors. However, model training requires high-quality curated datasets to ensure robust generalization to unseen datasets. Hence, we developed an artificial intelligence (AI) guided approach to assist with radiologist tumor delineation of partially segmented computed tomography datasets containing primary (adnexa) tumors and metastatic (omental) implants. AI guidance was implemented by training a 2D multiple resolution residual network trained with a dataset of 245 contrast-enhanced CTs with partially segmented examples. The same dataset curated through AI guidance was then used to refine two pretrained transformer models called SMIT and Swin UNETR. The models were independently tested on 71 publicly available multi-institutional 3D CT datasets. Segmentation accuracy was computed using the Dice similarity coefficient metric (DSC), average symmetric surface distance (ASSD), and the relative volume difference (RVD) metrics. Radiomic features reproducibility was assessed using the concordance correlation coefficient (CCC). Training with AI-guided segmentations significantly improved the accuracy of both SMIT (p $=$ 6.2$e^{-5}$) and Swin UNETR (p $=$ 2$e^{-4}$) models compared with using a partially delineated training dataset. Furthermore, SMIT-generated segmentations resulted in more reproducible features compared to Swin UNETR under multiple feature categories. Our results show that AI-guided data curation provides a more efficient approach to train AI models and that AI-generated segmentations can provide reproducible radiomics features.
\end{abstract}

\section{Introduction}

High grade serous ovarian cancer (HGSOC) accounts for most ovarian cancer-related deaths~\cite**{siegel2024cancer}. The prognosis for this disease is poor due to high cancer recurrence and presentation of high-volume multi-site disease. Multi-region analysis (primary tumor and peritoneal implants) is essential to accurately capture tumor diversity and identify key determinants of prognosis and treatment response to ultimately guide the selection of efficacious treatments to prolong life on treatment~\cite**{veeraraghavan2020integrated,crispin2023integrated}. However, multi-region radiomics analysis is impractical for clinical use due to the need for detailed manual tumor segmentation, which is labor-intensive. Hence, the goal of this work was to develop automated segmentation methods to extract multi-regional omental implants occurring in the peritoneum with primary disease occurring in the adnexa together.  

Deep learning has demonstrated success in generating automated tumor segmentation involving various disease sites including the brain~\cite**{hsu2021automatic}, lung~\cite**{jiang2021unpaired,zhang2022cross}, adrenal masses~\cite**{robinson2022machine}, pancreas~\cite**{li2021dual,chen2023pancreatic}, as well as colorectal metastases~\cite**{zhang2023automatic}. However, models are typically restricted to tumors occurring in specific organs. Prior works have undertaken the segmentation of ovarian cancer metastases, such as level set-based region growing applied to tumors with distinct appearance from background tissue such as hypodense perihepatic and perisplenic metastases~\cite**{liu2014tumor} as well as nnU-Net~\cite**{isensee2021nnu} used for more challenging tumors in the adnexa and omental implants~\cite**{buddenkotte2023deep}. However, the high intra-tumor heterogeneity and reduced contrast on CT resulted in relatively low accuracies. 

Our approach extends these methods by employing state-of-the-art transformers~\cite**{dosovitskiy2020vit}, which are more effective in capturing the long-range spatial context within images, thus capable of higher accuracy than convolutional neural networks. In particular, our architecture combines a transformer encoder with a convolutional U-Net style decoder with skip connections, thus combining the benefits of transformer and convolutional networks~\cite**{chen2024transunet}. Furthermore, the transformer encoder is pretrained as a foundation model, with a large number of diverse CT datasets (n $=$ 10,412 scans) using self-supervised learning that learn to extract features easily transferable to segmentation tasks~\cite**{jiang2022self}. 

A key challenge in training models to segment ovarian cancers is the difficulty in providing well-curated labeled datasets that contain accurate delineation of islands of tumors dispersed across the peritoneum. Clinically, it is more practical to delineate the primary adnexal tumors and one or few isolated islands of omental implants instead of all implants or one or few slices of the omental implants. This results in partial labels that can be challenging to be utilized directly for training a segmentation model. Partial labels have been combined with pseudo label generation with AI models and iterative retraining for various segmentation tasks typically involving organs~\cite**{wang2022semi,rangnekar2023semantic,basak2023pseudo,wu2023compete}. In this work, we adopt a `clinician in the loop' AI-guided approach, whereby the initial set of partial delineations is used to train a secondary AI model, which then provides AI guidance or completed segmentations that can be verified and edited as needed by the clinician. Our AI-guided approach reduces the time needed to carefully produce manual delineation of every tumor while providing good quality labels that improve the accuracy of the segmentation models. We will be releasing all of the model weights and inference codes for reproducibility post acceptance. 

\section{Method}

\subsection{Study Design, Patients, and Controls}
\label{sec:waiver}
This retrospective analysis was approved by the local institutional review board, with a waiver for written informed consent, and was compliant with the Health and Insurance Portability and Accountability Act. The public domain dataset from open source The Cancer Imaging Archive (TCIA) was used compliant with TCIA policies, such as maintaining participants’ privacy, accessing the data securely, and publication guidelines. 

\subsection{Datasets}
\label{sec:datasets}

\textbf{The training dataset} consists of 245 institutional 3D CECT scans of patients diagnosed with Stage II to IV HGSOC~\cite**{boehm2022multimodal}. These scans were acquired using different scanner manufacturers including GE (n $=$ 170) and the rest (n $=$ 75) from a mix of Siemens, Philips, and Toshiba scanners. A majority of the CECT scans were acquired at 120 KVp (range of 80 KVp to 140 KVp) and reconstructed with standard convolution kernels (5 mm thickness; range of 2.5 mm to 5 mm). The dataset contained fully annotated adnexal tumor masses and partially annotated omental implants. Partial annotations refer to cases where delineations were made to include either part of omental implants or some islands. Partial delineations of the omental implants were improved upon as discussed in Section~\ref{sec:aiguidedapproach}. Segmentation models were trained with 75\% (n $=$ 183) of the data and validated on 25\% (n $=$ 62 scans) to select the best model for independent testing. 

The \textbf{independent testing dataset} consisted of publicly available multi-institutional ovarian-TCIA dataset (n $=$ 71)~\cite**{vargas2017radiogenomics}. Precise radiologist-generated segmentations of all visible adnexal masses and omental implants were used for evaluations. The summary statistics for all the datasets used in this study are provided in Table~\ref{tab:datastatistics}.

\begin{table}[t]
\centering
\caption{Summary statistics of the training, validation, and test datasets used in this study. Median values are reported with inter-quartile ranges (IQR) where applicable.}
\label{tab:datastatistics}
\def\arraystretch{1.25}

\resizebox{\textwidth}{!}{%
\begin{tabular}{llll}
\textbf{} & Training (Institutional) & Validation (Institutional) & Test/Evaluation (Public) \\
\shline
Number of images & 19,995 & 5,844 & 9,300 \\
Number of scans & 183 & 62 & 71 \\
Volume - Adnexal tumors (cc) & 32.33 [6.32, 723.54] & 58.88 [3.37, 750.32] & 129.73 [35.47, 323.07] \\
Volume - Omentum implants (cc) & 52.62 [9.39, 392.70] & 43.65 [6.21, 200.01] & 37.90 [1.87, 200.1] \\
Axial-plane resolution (mm) & 0.80 [0.72, 0.86] & 0.79 [0.75, 0.87] & 0.76 [0.70, 0.82] \\
Voxel Spacing (mm) & 2.5 – 5.0 & 2.5 – 5.0 & 2.0 – 10.0 \\
Slice thickness (mm) & 5.0 [2.5, 5.0] & 5.0 [2.5, 5.0] & 5.0 [1.0, 10.0] \\
KVp (mA) & 120 [80, 140] & 120 [100, 130] & 120 [100, 140] \\
\bottomrule
\end{tabular}%
}
\end{table}

\begin{figure}[t]
    \centering
    \includegraphics[width=0.95\linewidth]{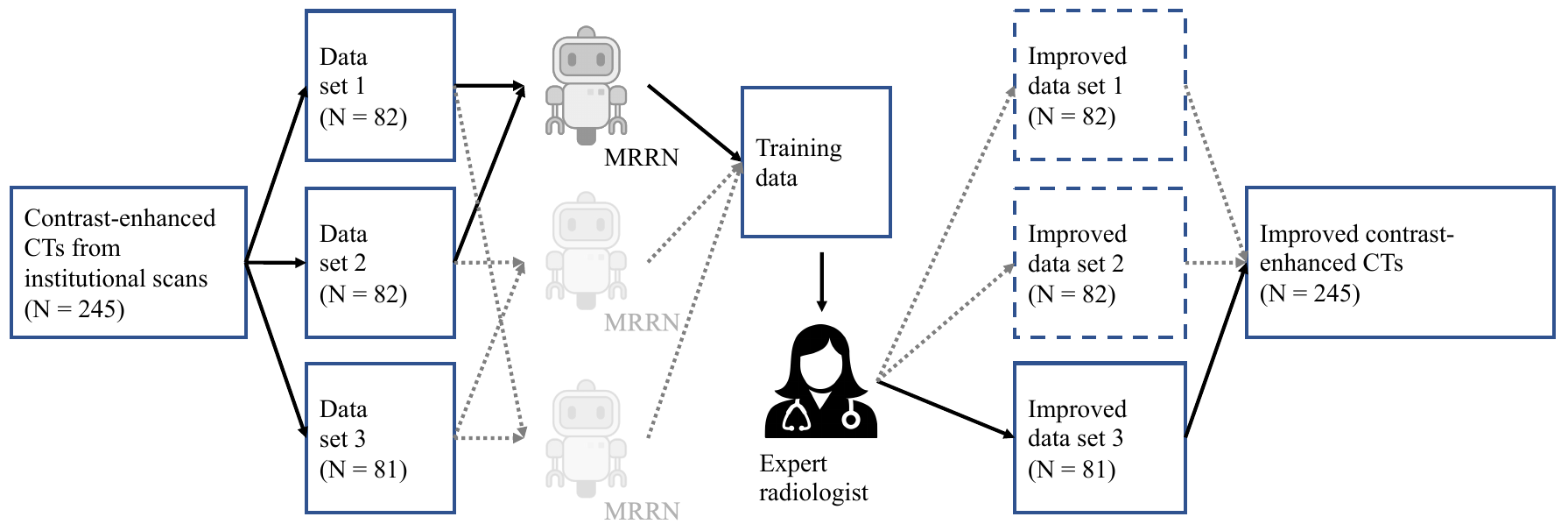}
    \caption{Pipeline for AI-guided labeling refinement in the CECT scans. Institutional scans (n $=$ 245) were split into three datasets for cross-validation. Each dataset fold was processed using the MRRN and further refined by expert radiologists, producing improved `AI-guided' manual delineations. The solid lines illustrate the workflow for one fold, while the dotted lines indicate the workflow for the other two folds.}
    \label{fig:mrrnaiguided_improvement}
\end{figure}

\subsection{AI-guided segmentation label refinement}
\label{sec:aiguidedapproach}

A three-step approach was used to refine manual delineations: (1) A 2D MRRN~\cite**{jiang2018multiple} was trained via 3-fold cross-validation on CT scans containing partially-delineated omental implants to generate AI model-based pseudo-segmentations. (2) Scans with low segmentation accuracy (Dice similarity coefficient $<$ 0.5) between pseudo-segmentations and manual delineations were flagged for expert review, and (3) Radiologist refined the segmentations as needed, erasing incorrect or false region segmentations with a single click. The AI-guided segmentation refinement approach is summarized in Figure~\ref{fig:mrrnaiguided_improvement}. The MRRN was trained with the combination of Dice and cross-entropy loss on cropped and Gaussian-smoothed images, in order to provide tumor-relevant spatial context with reduced image noise for segmentation. AI-guided editing was performed in one iteration without additional refinements to the AI-guidance model.

\subsection{Transformer model architectures}
\label{sec:foundationmodels}

\begin{figure}[t]
    \centering
    \includegraphics[width=0.95\linewidth]{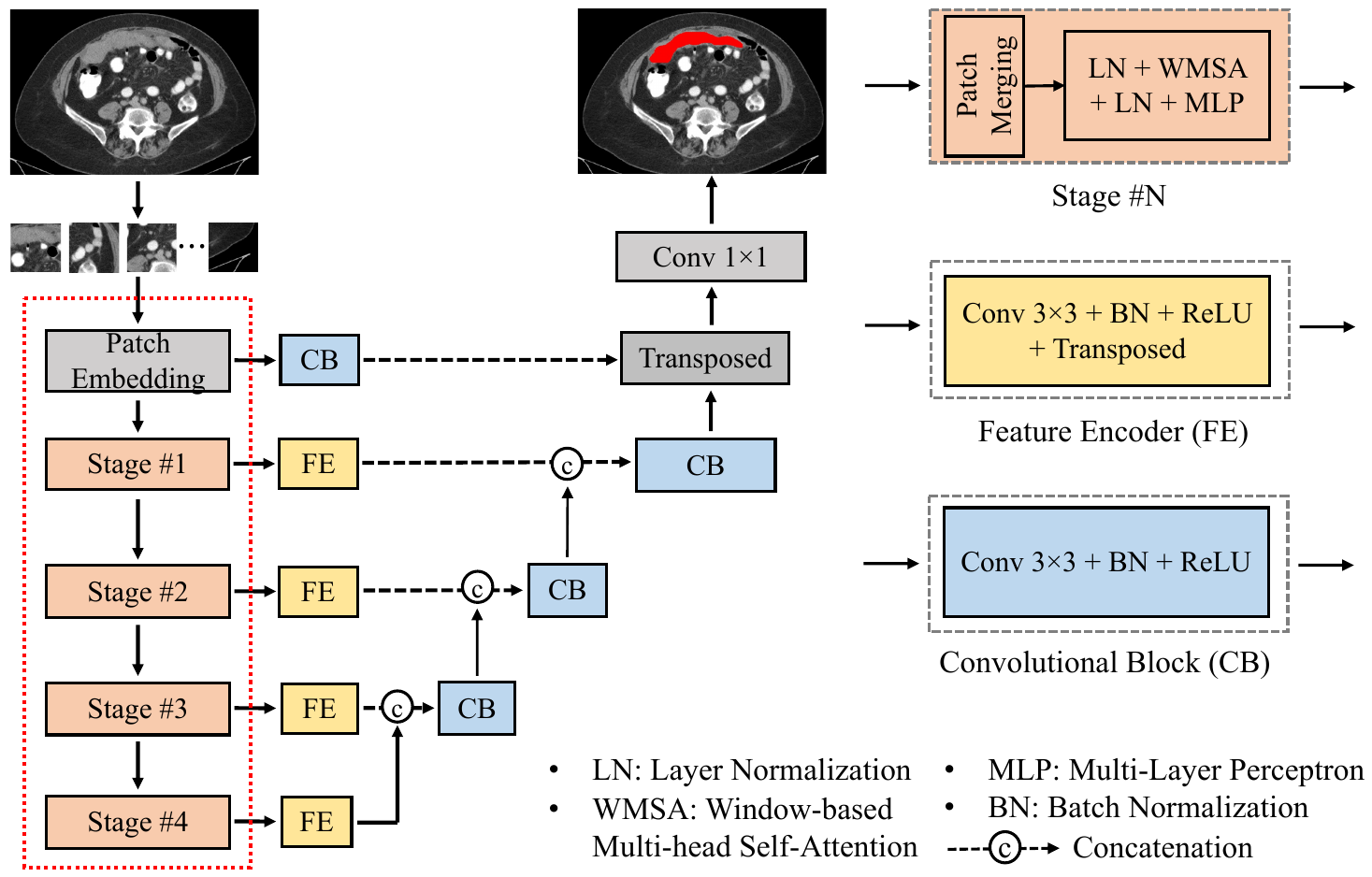}
    \caption{Architecture schematic of the 3D Swin transformer encoder combined with a U-Net decoder used for the segmentation task. The dotted red box indicates the encoder pretrained on large-scale datasets and fine-tuned for this study. This example demonstrates the model's segmentation of omental implants, shown on a 2D slice for clarity and brevity.}
    \label{fig:swintx_segmentation}
\end{figure}

Figure~\ref{fig:swintx_segmentation} shows the schematic of the transformer foundation models used in this study. The SMIT model~\cite**{jiang2022self}, originally developed for segmenting abdominal organs, was refined with fine-tuning and compared with the Swin UNETR~\cite**{tang2022self} model. Both architectures utilized a hierarchical Swin transformer backbone~\cite**{liu2021swin} to encode the 3D features and a 3D U-Net decoder~\cite**{ronneberger2015u} to generate tumor segmentations. SMIT used a $2-2-8-2$ block configuration with additional windowed attention blocks implemented in the third stage, while Swin UNETR used a $2-2-2-2$ block configuration. The two networks also differed in their pretraining approach. SMIT employed a self-distillation framework that combined masked image prediction with token distillation tasks. Swin UNETR used pretext tasks such as masked image inpainting, rotation prediction, and contrastive coding. Of note, 3D scan volumes were masked and predicted as 3D image volumes.

Both models were fine-tuned on identical datasets using a combination of Dice and cross-entropy losses to segment tumors in the adnexa and omental implants as distinct tumor classes. The pretrained encoders provided from publicly available model configurations and weights were used to initialize the networks from the respective GitHub repositories. The U-Net decoder used for segmentation was initialized with random weights. Training used PyTorch~\cite**{paszke2019pytorch} and MONAI~\cite**{cardoso2022monai} on 4 $\times$ Nvidia A100 80GB GPUs, and voxel patch dimensions of 128 $\times$ 128 $\times$ 128 for SMIT and 96 $\times$ 96 $\times$ 96 for Swin UNETR, consistent with their model configurations, at spacing of 1.0 $\times$ 1.0 $\times$ 1.0 $mm^3$. Sliding window inference with 50\% Gaussian overlap was applied for full 3D CECT scan segmentation.

\subsection{Training curricula}
SMIT and Swin UNETR architectures were trained under multiple curricula to analyze the performance improvements. Specifically, both models were trained using the partially annotated dataset of omental implants (referred to as the \textit{partial} curriculum and with the radiologist-refined segmentations based on AI-generated pseudo segmentations (referred to as \textit{AI-guided} curriculum. Standard data augmentations $-$ including rotation, translation, and scaling, were applied using the MONAI framework~\cite**{cardoso2022monai}. Additionally, the AI-guided curriculum was also trained using advanced data augmentation techniques introduced in nnU-Net~\cite**{isensee2021nnu}, namely Gaussian noise, brightness adjustment, contrast adjustment, and gamma correction, which were performed on the fly during training. This enhanced setup is denoted as `+ nnU-Aug' in this paper. 

The adnexal tumors were fully delineated by radiologists and this training curriculum is referred to as `Full' and the models trained using `Full' with nnU-Aug are called `Full $+$ nnU-Aug' throughout the paper.

\subsection{Evaluation metrics and statistical analysis}
\label{sec:metrics}

Segmentation accuracy was computed by comparing the tumor segmentations to manual delineations using DSC, average symmetric surface distance (ASSD), and relative volume difference (RVD) metrics. Bland-Altman plots computed the differences in the segmented versus manually delineated tumor volumes to supplement the analysis. Statistical comparisons measured the performance improvements of the individual models trained with only partial delineations versus AI-guided delineations versus `+nnU-Aug' training instances using paired two-sided Wilcoxon signed-rank tests at 95\% confidence level, $p <$ 0.05 considered significant. 
Manual editing effort for AI (MRRN pseudo segmentations)-guided manual delineations were measured using the added path length (APL) metric~\cite**{vaassen2020evaluation}.

\begin{figure}[t]
    \centering
    \includegraphics[width=0.98\linewidth]{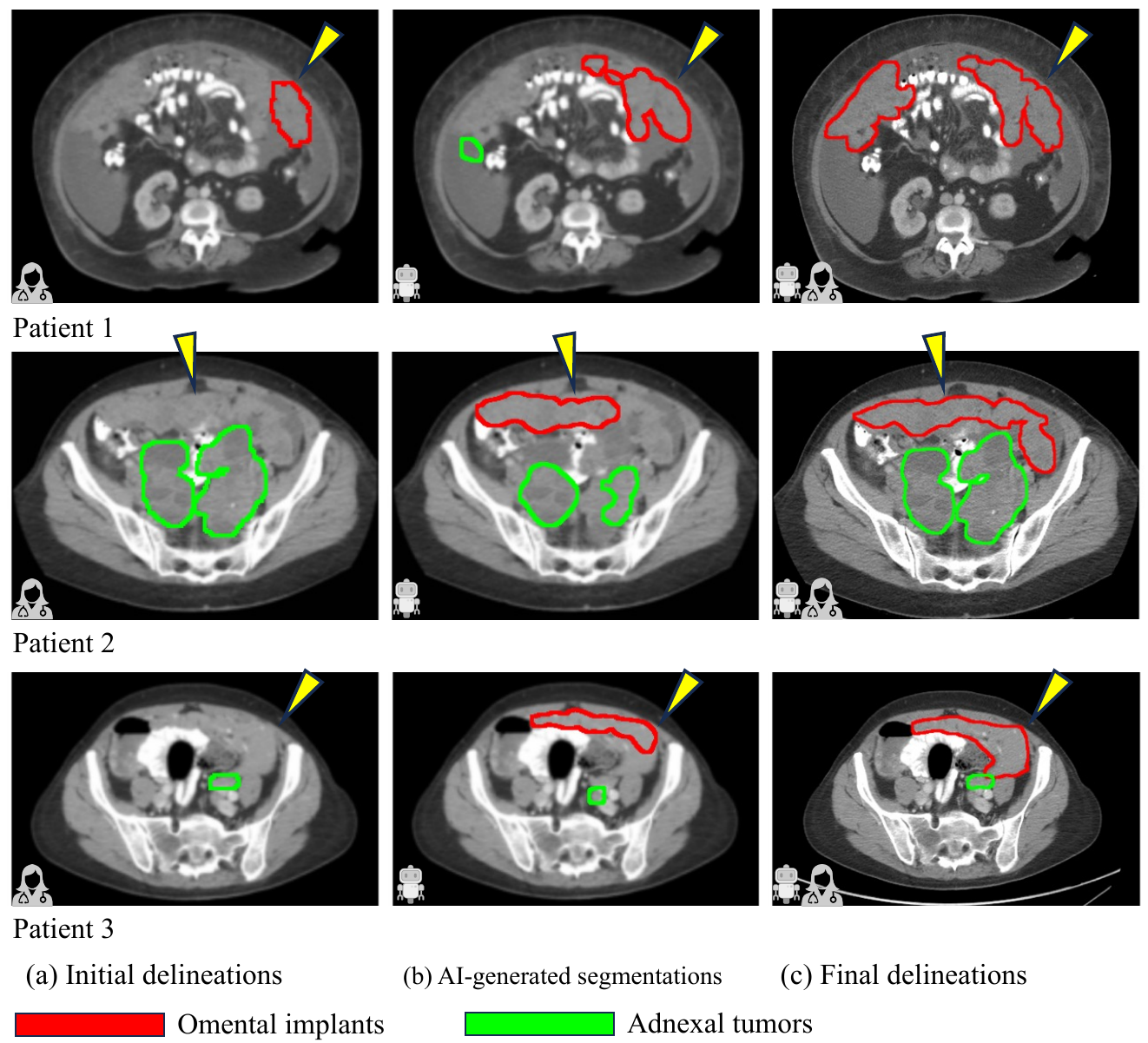}
    \caption{Representative cases illustrating the delineation refinement process of the metastatic lesions in the omentum with (a) initial partial delineations, (b) the AI-generated segmentations, and (c) the final radiologist delineations. Red and green contours represent the omental implants and the adnexal tumors respectively. The yellow pointers highlight the key regions where the refinements occurred. Note that the adnexal tumors were not further refined as they were fully delineated from the beginning itself.}
    \label{fig:aiguidedimprov_omentum_gts}
\end{figure}

Finally, multiple radiomics features (n $=$ 293) quantifying the intra-tumor texture heterogeneity were extracted using the CERR framework~\cite**{deasy2003cerr,apte2018extension}, compliant with the Image Biomarkers Standardization Initiative (ISBI)~\cite**{zwanenburg2020image} to analyze the similarity of radiomic features extracted within AI models' versus their nnU-Aug segmentations. Radiomic features for omental implants that included multiple isolated lesions were averaged to extract a single scalar value for each radiomic feature. Analyzed radiomic features included shape (n $=$ 23), first-order or intensity histogram features (n $=$ 22), gray level co-occurrence matrix (GLCM n $=$ 130), gray level run length matrix (GLRLM n $=$ 80), gray level site zone matrix (GLSZM n $=$ 16), gray level dependence matrix (GLDM n $=$ 17), and gray tone difference matrix (GTDM n $=$ 5). Default parameter settings included kernel size of 3 $\times$ 3 $\times$ 3 for textures, images resampled to 1 $mm$ on all dimensions, and intensity clipping [-175, 250] in the CERR pipeline. 

Unlike most existing works that considered the segmentation of single tumors of a fixed number of targets, our work concerns the segmentation of a variable number of lesions as the omental implants are dispersed across the peritoneum, ranging from one to many lesions. Hence, image-wide computation of accuracy such as image-level DSC over-simplifies and over-estimates the accuracy. Specifically, accuracy in our case combines both detection of individual lesions and their segmentation. For this purpose, we computed lesion-level accuracies, whereby individual segmented lesions were separated into distinct regions or clusters by morphological processing using connected components extraction. Next, corresponding AI and manually segmented lesions were identified by computing spatial overlap and deeming those with at least $\geq$ 0.5 as corresponding lesions for accuracy computation. Such an overlap threshold was deemed to be a detection threshold in prior work on segmentation and was used as is~\cite**{bilic2023liver}. Precision and recall metrics were also computed to measure lesion-level detection accuracies.

\begin{table}[t]
\centering
\caption{Comparison of the testing accuracy metrics (Dice similarity coefficient - DSC, average symmetric surface distance - ASSD, and relative volume difference - RVD) for the two models and different training curricula on segmenting omental implants and adnexal tumors.}
\label{tab:mainresults}
\def\arraystretch{1.25}
\resizebox{\textwidth}{!}{%
\begin{tabular}{l|llll|llll}
Model & \multicolumn{4}{l}{Omental implants} & \multicolumn{4}{l}{Adnexal tumors} \\
 & Curriculum & DSC ($\uparrow$) & ASSD (mm $\downarrow$) & RVD ($\downarrow$) & Curriculum & DSC ($\uparrow$) & ASSD (mm $\downarrow$) & RVD ($\downarrow$)\\ \shline
SMIT & Partial & 0.73 $\pm$ 0.09 & 2.64 $\pm$ 1.71 & 0.22 $\pm$ 0.29 &  &  &  &  \\
 & AI-guided & 0.76 $\pm$ 0.12 & 2.12 $\pm$ 2.29 & 0.07 $\pm$ 0.29 & Full & 0.81 $\pm$ 0.08 & 2.63 $\pm$ 1.74 & 0.19 $\pm$ 0.25 \\
 & \hspace{0.5em}+ nnU-Aug & 0.77 $\pm$ 0.10 & 2.90 $\pm$ 3.05 & 0.10 $\pm$ 0.27 & \hspace{0.5em}+ nnU-Aug & 0.83 $\pm$ 0.07 & 2.36 $\pm$ 1.57 & 0.17 $\pm$ 0.20 \\
 \midrule
\begin{tabular}[c]{@{}l@{}}Swin UNETR\\ \end{tabular} & Partial & 0.73 $\pm$ 0.10 & 3.54 $\pm$ 1.95 & 0.25 $\pm$ 0.31 &  &  &  &  \\
 & AI-guided & 0.76 $\pm$ 0.11 & 3.22 $\pm$ 4.39 & 0.12 $\pm$ 0.29 & Full & 0.77 $\pm$ 0.16 & 3.56 $\pm$ 4.66 & 0.10 $\pm$ 0.31 \\
 & \hspace{0.5em}+ nnU-Aug & 0.76 $\pm$ 0.11 & 3.48 $\pm$ 6.80 & 0.15 $\pm$ 0.30 & \hspace{0.5em}+ nnU-Aug & 0.79 $\pm$ 0.12 & 3.31 $\pm$ 2.32 & 0.09 $\pm$ 0.29 \\
 \bottomrule
\end{tabular}%
}
\end{table}

\section{Results}

\subsection{Increasing efficiency of manual delineations with AI-guiance}
We measured the radiologist edits when using AI guidance for completing partial delineations. Seventy two out of 245 scans were identified to have large deviations in the segmentation of the omental implants using the MRRN method compared to radiologist edits. The remaining with high similarity of MRRN segmentations with radiologist preliminary delineations were visually validated and found to not need additional editing. AI guidance using MRRN reduced the number of voxels requiring radiologist editing as shown by a lower APL median of 39.55 and IQR of 10.41 to 63.13 when compared to radiologist delineations on the partial delineations (APL median of 106.49, IQR of 55.22 to 208.44). Figure~\ref{fig:aiguidedimprov_omentum_gts} shows representative examples with initial partial segmentations with missing segmentations of omental implants, the AI-generated segmentation, and the radiologist edits performed on the AI segmentations. 

\begin{figure}[t]
    \centering
    \includegraphics[width=0.98\linewidth]{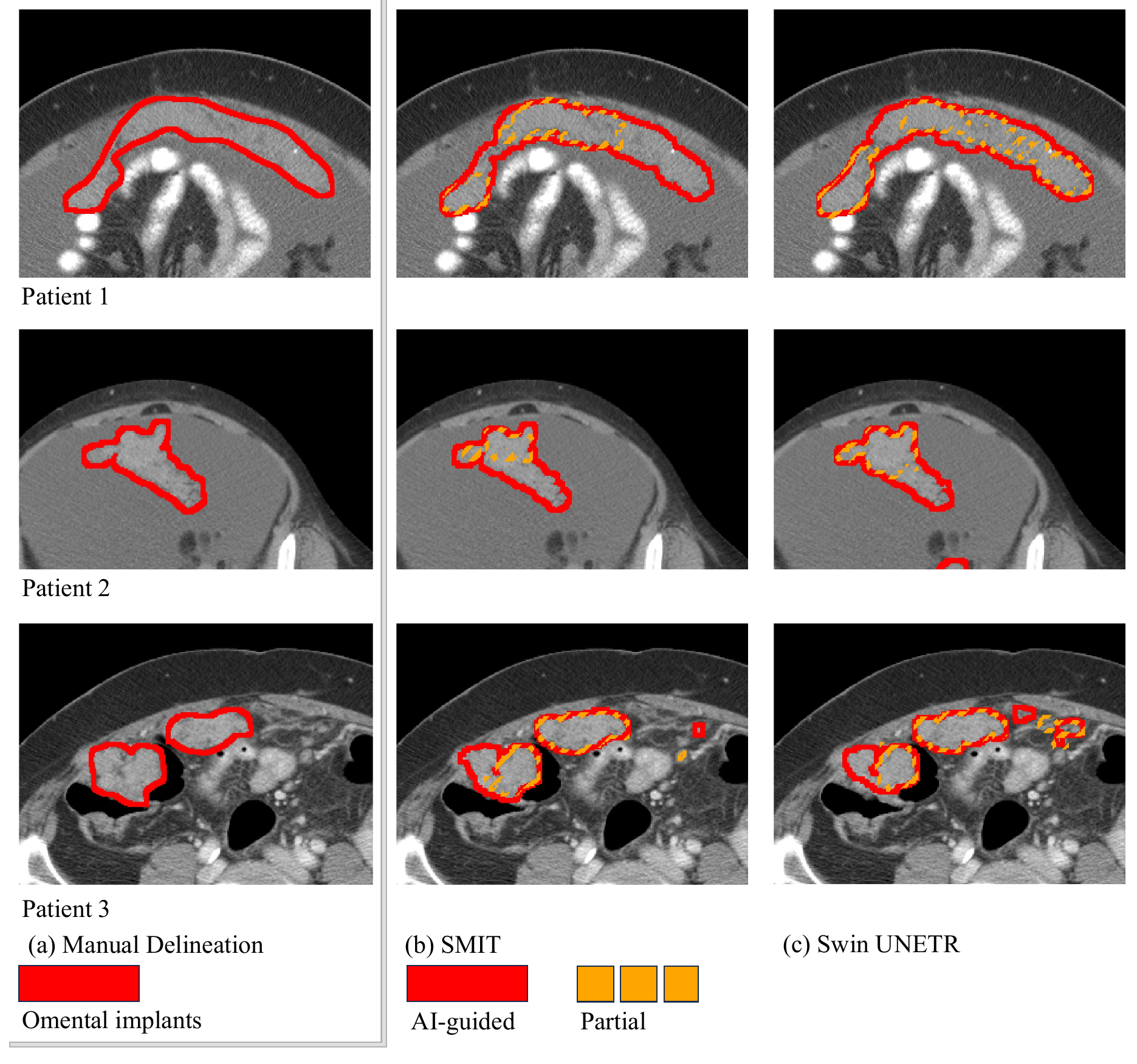}
    \caption{Segmentations of omental implants using the AI models trained with different curricula. (a) shows the manual delineations, and (b) and (c) shows the results of the SMIT and Swin UNETR models. The solid red lines highlight the AI-guided curriculum results, while the dashed orange lines indicate the partial curriculum results with visible improvements.}
    \label{fig:aiguidedimprov_omentum_ops}
\end{figure}

\subsection{Benefit of AI-guided segmentations on models' accuracy}

\begin{figure}[t]
    \centering
    \includegraphics[width=0.98\linewidth]{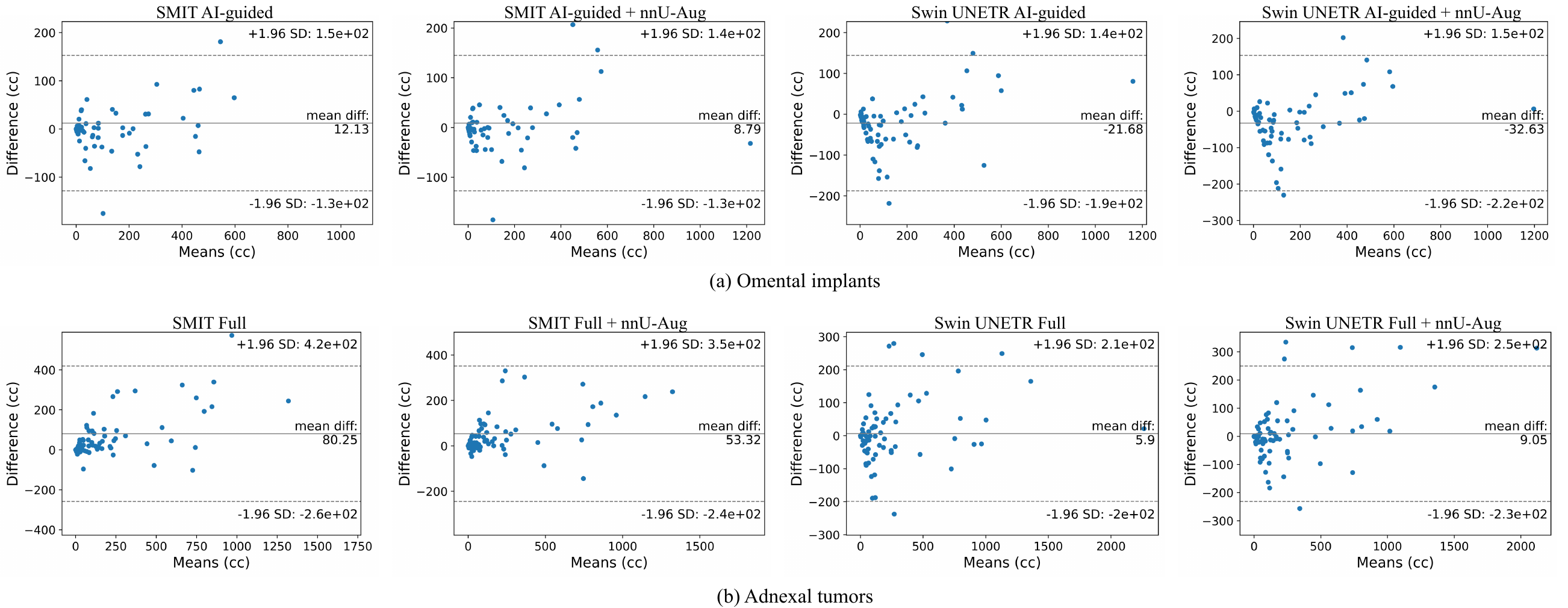}
    \caption{Bland-Altman plots showing tumor volumes generated by the multiple AI models in comparison to manual delineations for omental implants (a) and adnexal tumors (b). The X-axis shows (manual volume + AI volume)/2 and the y-axis depicts (manual volume - AI volume)/2.}
    \label{fig:blandaltman}
\end{figure}

Figure~\ref{fig:aiguidedimprov_omentum_ops} shows performance improvements in segmentations on the test set when trained using AI-guided segmentations. As shown, both models produced significantly more accurate segmentation of the omental implants when trained using AI-guided radiologist delineations compared to training with partially labeled cohorts (SMIT p $<$ 0.0001, Swin UNETR p $=$ 0.0002). Precision (SMIT p $=$ 0.04, Swin UNETR p $=$ 0.020) and recall (SMIT p $=$ 0.04, Swin UNETR p $=$ 0.038) also improved for detecting omental implants. Further analysis of differences in segmented tumor volumes using Bland Altman plots showed that Swin UNETR resulted in over-segmentations compared to SMIT (Figure~\ref{fig:blandaltman}). SMIT provided results that closely followed manual delineations. 

\begin{figure}[t]
    \centering
    \includegraphics[width=0.98\linewidth]{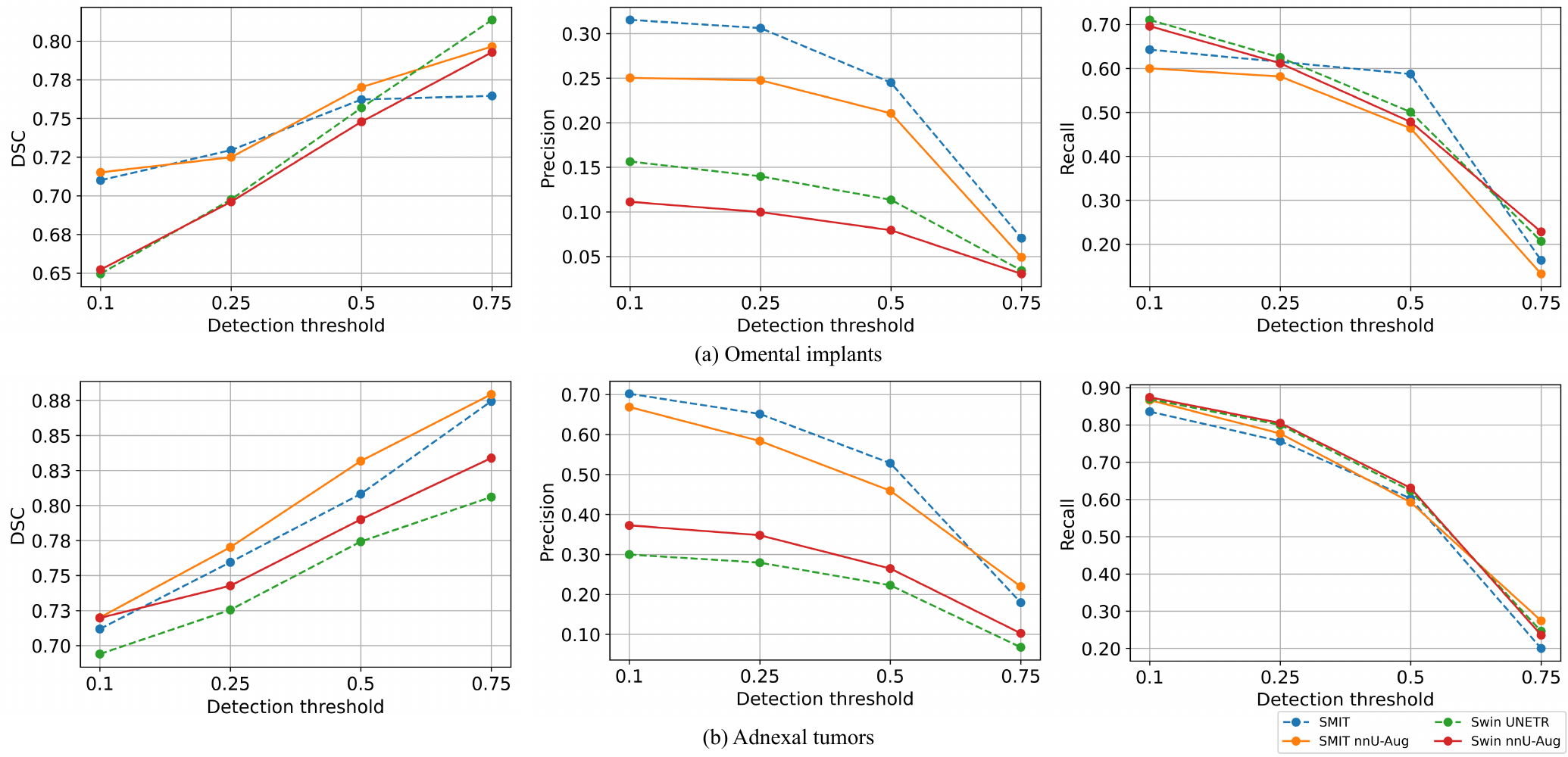}
    \caption{Calibration curves for omental implants (a) and adnexal tumors (b), showing the performance of the SMIT and Swin UNETR models under varying detection thresholds. The columns represent the Dice similarity coefficient (DSC), precision, and recall metrics. Solid lines indicate the use of the nnU-Aug curriculum, while dashed lines indicate the performance of models trained with the standard curriculum.}
    \label{fig:calibrationcurves}
\end{figure}

Analysis of the impact of tumor lesion detection threshold on the metrics showed that SMIT achieved consistently higher DSC and precision, but similar recall as Swin UNETR for both tumor types (Figure~\ref{fig:calibrationcurves}). Additionally, we found that the tumor lesion detection threshold of 0.5 balanced precision and recall rates across all models in our dataset. Hence, we use it as the default threshold for reporting all the accuracies.

\subsection{Benefit of additional augmentation methods on accuracy}

\begin{figure}[t]
    \centering
    \includegraphics[width=0.98\linewidth]{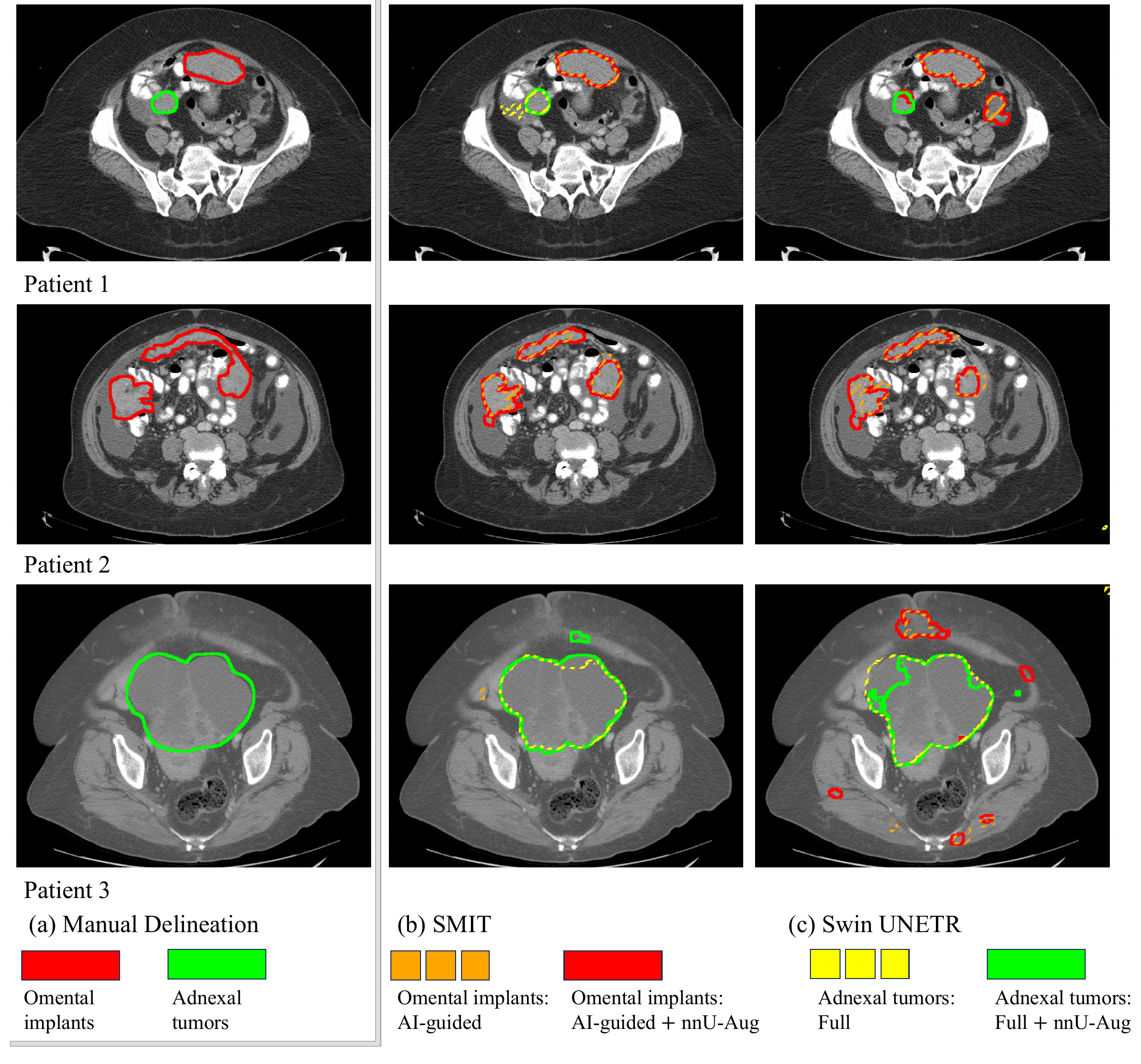}
    \caption{Segmentations for omental implants and adnexal tumors on three representative cases by the SMIT (b) and Swin UNETR (c) models, alongside the expert delineation (a). Corresponding legends for the curricula are provided, with a focus on outputs using the `nnU-Aug' curriculum highlighted with solid lines.}
    \label{fig:resultingfigures}
\end{figure}

Accuracy gains increased from 0.81 to 0.83 for the adnexal tumors in the case of SMIT and from 0.77 to 0.79 in the case of Swin UNETR (Table~\ref{tab:mainresults}). However, nnU-Aug-based improvements were not statistically significant when compared to standard augmentations for both lesion types (SMIT: omental implants p $=$ 0.11, adnexal tumors p $=$ 0.24; Swin UNETR: omental implants p $=$ 0.18, adnexal tumors p $=$ 0.19). Figure~\ref{fig:resultingfigures} shows example instances showcasing the effect of nnU-Aug on both the models evaluated in this paper.

\subsection{Reproducibility of radiomic features}

\begin{figure}[t]
    \centering
    \includegraphics[width=0.98\linewidth]{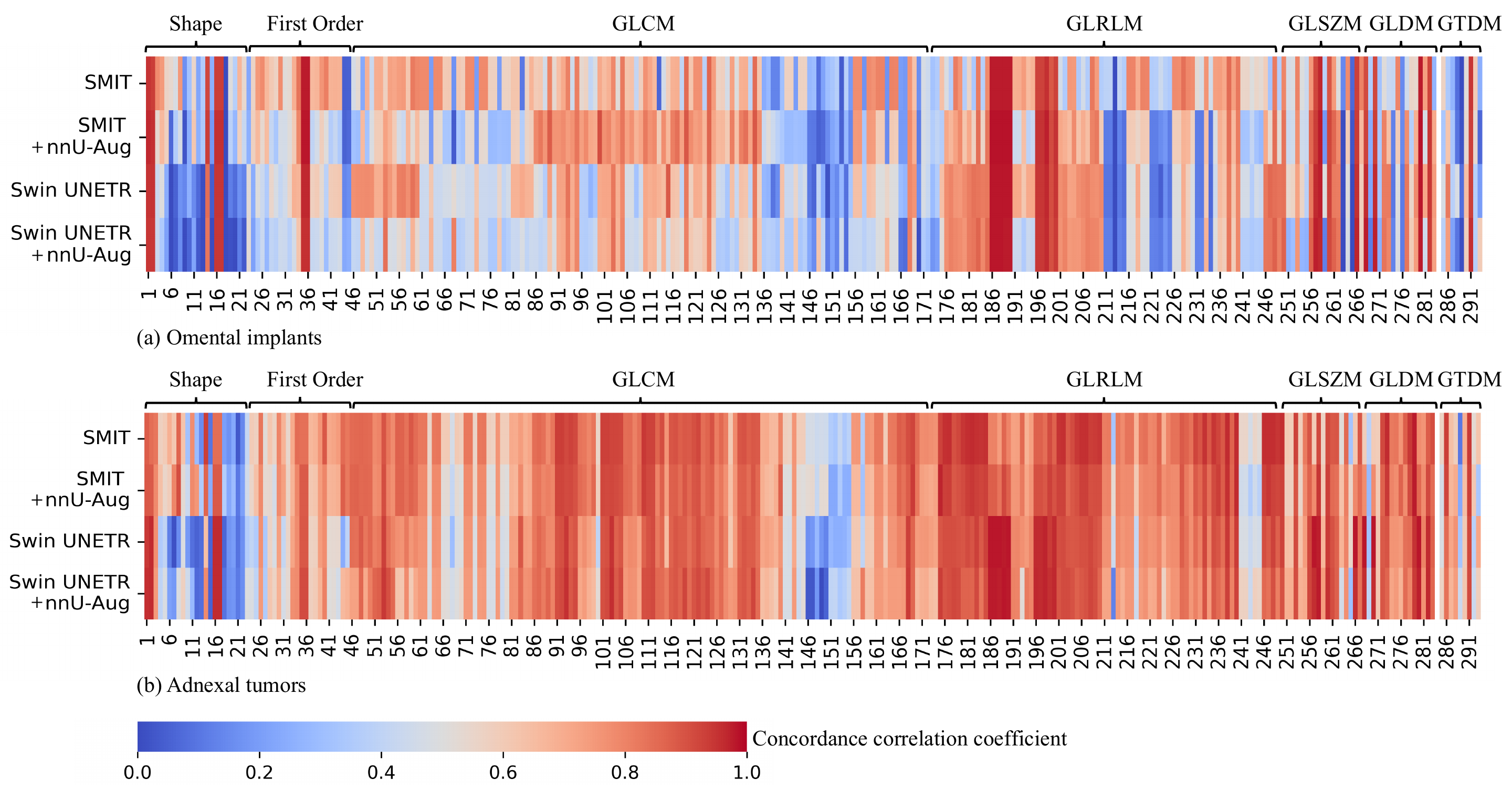}
    \caption{Concordance correlation Coefficient (CCC) on the radiomics features for omental implants (a) and adnexal tumors (b) between manual delineations and auto-segmentations.}
    \label{fig:radiomics}
\end{figure}

Feature similarity was computed using concordance correlation coefficient (CCC) for all the 293 features and averaged for the individual feature categories for presentation as shown in Table~\ref{tab:radiomics}. As shown, SMIT and SMIT with nnU-Aug methods resulted in higher reproducibility of radiomic feature categories for both omental implants and tumors in the adnexa. nnU-Aug applied to SMIT increased reproducibility of GLCM, GLRLM, and features not corresponding to shape and first-order features for the omental implants, whereas SMIT resulted in higher reproducibility for first-order, GLCM, and GLRLM features computed from within the adnexa tumors. Feature-specific reproducibility analysis is presented in Figure~\ref{fig:radiomics}, which shows higher reproducibility for tumors in the adnexa using both SMIT and Swin UNETR methods across multiple feature categories, but higher reproducibility with SMIT for specific categories in the case of omental implants.  

\begin{table}[t]
\centering
\caption{Percentage of reproducible features with CCC $\geq$ 0.75 across different radiomic feature categories for omental implants and adnexal tumors, evaluated for the different curricula.}
\label{tab:radiomics}
\resizebox{\textwidth}{!}{%
\begin{tabular}{l|lllll|lllll}
Method & \multicolumn{5}{l}{Omental implants reproducible features \% ($\uparrow$)} & \multicolumn{5}{l}{Adnexal tumors reproducible features \% ($\uparrow$)} \\
 & \multicolumn{1}{l}{Shape} & First-order & GLCM & GLRLM & Rest & Shape & First-order & GLCM & GLRLM & Rest \\ \shline
  SMIT & \textbf{30.43}& \textbf{31.82}& 23.08& 38.75& 33.33& 34.78& \textbf{36.36}& \textbf{69.23}& \textbf{78.75}& 53.85\\
\hspace{0.5em}+ nnU-Aug & 21.74& 13.64& \textbf{25.38}& \textbf{26.25}& \textbf{35.90}& \textbf{38.09} & 31.82& 63.85& 75.00& \textbf{64.10}\\
Swin UNETR & 21.74& 9.090& 16.92 & 43.75& 43.59& 21.74& 22.73& 56.23& 83.75& 64.20\\
\hspace{0.5em}+ nnU-Aug & 21.74& 9.090& 7.690& 46.25& 38.46& 21.74& 22.73& 46.15& 72.50& 56.41\\
\bottomrule
\end{tabular}%
}
\end{table}

\section{Discussion}

This work focused on using an AI-guided radiologist segmentation approach to improve the performance of deep learning methods. Two transformer-based models that were pretrained on a large number of 3D CT scans unrelated to the datasets were used to initialize the models in order to reduce the need for labeled ovarian cancer datasets. Our analysis showed significant accuracy improvements for omental implants with dispersed lesions occurring within the peritoneum when using an AI-guided radiologist segmentation approach irrespective of the AI model used. Furthermore, our approach also demonstrated that AI-generated pseudo segmentations to guide radiologist delineations reduced radiologist effort in delineating dispersed omental implants. These results are in line with prior efforts that showed a reduction in manual effort in delineation when using AI methods~\cite**{cha2021clinical,goel2022deployed}. These results indicate that AI-guided methods can be used to reduce radiologists' efforts in providing well-curated datasets while also improving the accuracy of further analysis to provide automated methods. 

Our approach differs in key aspects from existing AI-guided methods in that, AI-guided refers to the generation of pseudo labels by AI models that are directly used for training and improving AI models by providing more numbers of labeled datasets. Our approach explicitly involves a `clinician-in-the-loop' approach to verify and edit the AI segmentations as required. Such an approach is essential when developing methods for automated segmentation of tumors for downstream analysis including quantifying tumor burden response of dispersed and isolated lesions to treatments as well as radiomics methods. 

This is also the first approach to use transformers in the encoder with a convolutional network used in the decoder to segment metastatic omental implants and primary adnexal tumors. Previously, an off-the-shelf nnU-Nnet method was refined and used to segment omental implants and adnexal tumors in~\cite**{buddenkotte2023deep}. However, the reported results showed poor performance for omental implants often due to the difficulty of automated segmentation for dispersed, irregularly shaped, and sparse deposits with poor soft-tissue contrast. Our results showed accuracy improvements by incorporating AI guidance, which partly alleviates the segmentation challenge by easing radiologist delineations. 

Limitations of this study include the lack of disaggregated analysis to assess performance differences across scanner manufacturers and acquisition locations, although generalization was partially addressed by evaluating a public dataset with images sourced from multiple institutions and different scanner manufacturers (The Cancer Imaging Archive). Additionally, the focus was limited to omentum implants and adnexal tumors due to the practical constraints in producing carefully curated radiologist delineations, leaving other metastatic sites unexplored. Expanding the segmentation task to include other ovarian metastatic sites, and exploring architectural improvements could further enable potential clinical applications for accurate assessment of overall disease volume.

\section{Conclusion}
This study demonstrated the utility of AI-guided radiologist segmentations improved the accuracy of transformer-based models to segment dispersed omental implants occurring in the peritoneum. In addition, the same models were also capable of simultaneously segmenting primary ovarian tumors occurring in the adnexa. Further study would extend the models to additional metastatic tumor sites and larger multi-institutional cohorts to enhance applicability in clinical and clinical trial studies. 

\ack
We thank the MSK MIND consortium and the NIH National Cancer Institute MSK Cancer Center Support Grant P30CA008748.

\section*{References}
\bibliography{references}
\bibliographystyle{dcu}

\end{document}